**Successive incommensurate spin orderings and excitations in multiferroic SrMnGe$_2$O$_6$**

Claire V. Colin,[1] Lei Ding,[2] Eric Ressouche,[3] Julien Robert,[1] Noriki Terada,[4] Frederic Gay,[1] Pascal Lejay,[1] Virginie Simonet,[1] Céline Darie,[1] Pierre Bordet[1], Sylvain Petit[5]

[1] *Université Grenoble Alpes, CNRS, Institut Néel, 38000, Grenoble, France*
[2] *ISIS Facility, Rutherford Appleton Laboratory, Harwell Oxford, Didcot OX11 0QX, United Kingdom*
[3] *Université Grenoble Alpes, CEA, IRIG, MEM, MDN, F-38000 Grenoble, France*
[4] *National Institute for Materials Science, Sengen 1-2-1, Tsukuba, Ibaraki 305-0047, Japan*
[5] *Laboratoire Léon Brillouin, Université Paris-Saclay, CNRS, CEA, CEA-Saclay, Gif-sur-Yvette F-91191, France*

**Abstract**

Anisotropic multiferroic properties of SrMnGe$_2$O$_6$ pyroxene single crystals were systematically investigated by means of magnetization, heat capacity, pyroelectric current measurement and elastic and inelastic neutron scattering experiments. Single crystal neutron diffraction allows us to unambiguously reveal the presence of two incommensurate magnetic orderings: a non-polar amplitude-modulated collinear sinusoidal magnetic structure emerges at T$_{N1}$=4.36(2)K followed by a polar elliptical cycloidal spin structure below T$_{N2}$=4.05(2)K. Pyroelectric current measurements on single crystal confirm the appearance of a spontaneous polarization within the (*ac*) plane below T$_{N2}$ associated with the latter magnetic symmetry through extended Dzyaloshinsky-Moriya mechanism. The magnetic phase diagram was calculated considering the three isotropic exchange couplings relevant in this system. The magnetic excitations spectra of SrMnGe$_2$O$_6$ measured by inelastic neutron scattering were successfully modeled using a set of exchange interactions consistent with this phase diagram.

**I introduction**

Strong coupling between magnetism and electricity in matter has become a central issue of condensed-matter physics from both fundamental and technological points of view. In so-called spin-driven multiferroics, an electric polarization emerges due to the symmetry breaking induced by the magnetic ordering. These last 15 years, a large variety of magnetic orderings was found to induce ferro- or ferrielectricity. Mainly three different microscopic models have been proposed to describe the observed ferroelectricity in different spin-driven ferroelectrics: for commensurate structures the relevant mechanism is exchange striction, whereas in incommensurate structures inverse Dzyaloshinskii-Moriya (DM) model[1,2] is proposed for cycloid-type ordering, while spin-dependent p-d hybridization model[3] is invoked in proper screw type of magnetic ordering. The inverse DM model arising from the antisymmetric spin-exchange interaction between canted spin sites has been successfully used to explain the emergence of ferroelectricity in multiferroïcs with cycloidal order (like prototypical multiferroïc $TbMnO_3$[4]) but also transverse-conical spin order (for instance in spinel type $CoCr_2O_4$[5]). In addition, Kaplan and Mahanti[6] have shown that the extended inverse DM effect in some specific systems contributes to microscopic electric polarization in both cycloid and proper-screw helical systems.

Competing antiferromagnetic exchanges are well known to lead to frustration and give rise to modulated magnetic phases and rich phase diagram. It is the case in Pyroxenes, which are historically of great importance in mineralogy and have gained recently interest in condensed matter physics because of the interplay between low dimensionality and magnetic frustration. The general formula of Pyroxenes is $AMX_2O_6$, where A is usually an alkali metal ion with a valence of 1+ (e.g. $Li^+$ and $Na^+$) or an alkaline earth ion with a valence of 2+ ($Mg^{2+}$, $Ca^{2+}$ and $Sr^{2+}$), M refers to trivalent or divalent transition metal ions (e.g. $Fe^{3+}$ and $Mn^{2+}$), and X represents $Si^{4+}$ or $Ge^{4+}$. Based on the observation of electric polarization under magnetic fields, Jodlauk et al [7] reported that the clinopyroxene (pyroxene with a monoclinic crystal structure) compounds $NaFeSi_2O_6$, $LiFeSi_2O_6$, and $LiCrSi_2O_6$ could display multiferroism. They suggested that the quasi-one-dimensional spin chain of $M^{3+}$ ions should be subject to spin frustration between intrachain and interchain interactions resulting in an incommensurate spin structure. These magnetic structures could induce ferroelectric polarization, possibly due to the formation of spiral spin ordering. Subsequent neutron diffraction and electric polarization studies revealed that not only $NaFeSi_2O_6$ [8–10] but also $NaFeGe_2O_6$[9,11–14] are spin driven multiferroics, while $LiFeSi_2O_6$[15,16], $LiCrSi_2O_6$[17–19] and also $CaMnGe_2O_6$[20] are linear magnetoelectric materials. In pyroxene triangular topology, the $3d^5$ electronic configuration of the transition metal is a crucial feature to trigger magnetic frustration because it is the only one that gives rise to predominant and uniform AFM interactions along the octahedra chains[21]. Recently, we have established that the divalent pyroxene $SrMnGe_2O_6$, with $Mn^{2+}$ in a $3d^5$ electronic configuration, present indeed multiferroism[22].

The structure of SrMnGe$_2$O$_6$, (*C2/c*, *a* = 10.346(2) Å, *b* = 9.420(2) Å, *c* = 5.511(1) Å, β = 104.669(5)) shown in figure 1(a), is characterized by zigzag spin chains of Mn$^{2+}$ running along the c-axis. Along the a-axis, the spin chains are connected via GeO$_4$ tetrahedra, and thus intrachain interaction (J) along the c-axis is strongest while the weaker interchain interactions (J$_1$ and J$_2$) can be a source of competing spin interactions. Therefore, the magnetic arrangement can be regarded as a spin-frustrated network of quasi-one-dimensional spin chains, thus prone to the formation of incommensurate spin ordering.

Recently, we have succeeded in growing large single crystal of SrMnGe$_2$O$_6$ by the floating zone furnace technique, as it can be observed in Figure 1(b). In the present study, we extend the investigation done previously on powders[22] by using these single crystals and access the anisotropic properties. Hereafter are reported our single-crystal magnetization, heat capacity, pyroelectric current measurements as well as elastic and inelastic neutron scattering experiments.

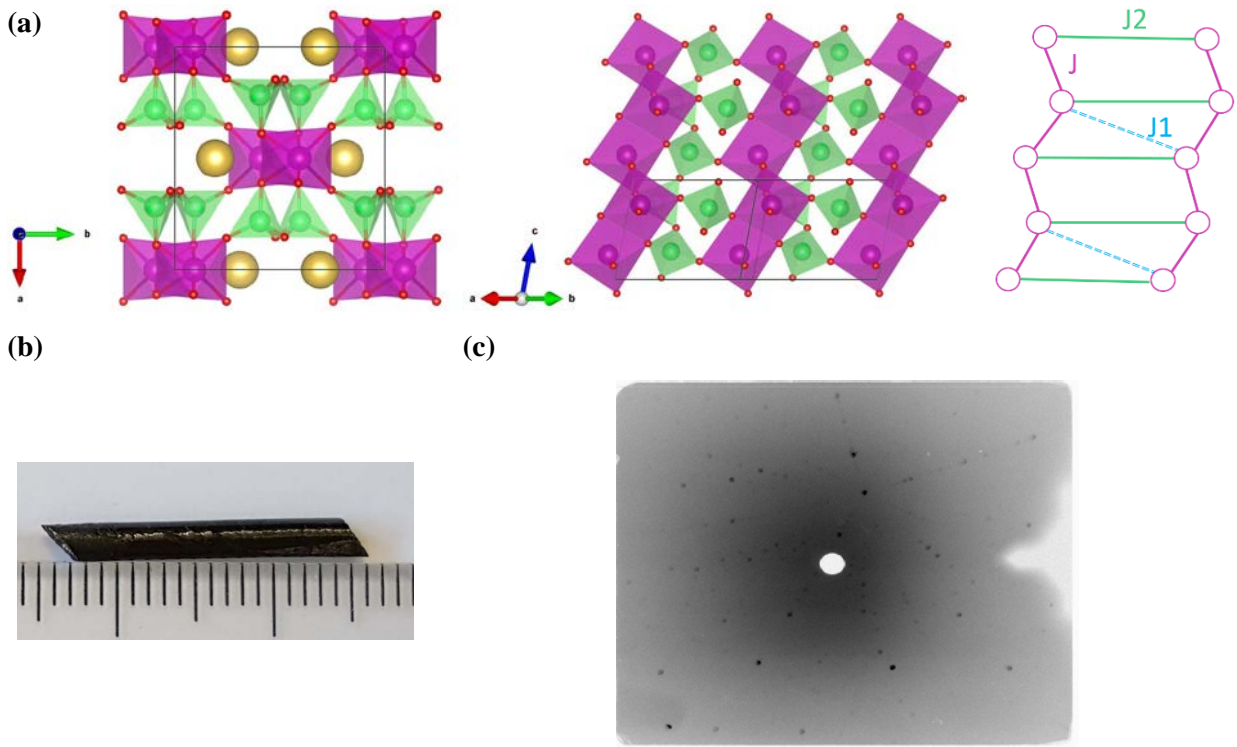

*Figure 1 (a) Crystal structure of SrMnGe$_2$O$_6$ (C2/c) projected along two different directions. Atoms are represented by colored spheres: Sr in gold, Mn in purple, Ge in green and O in red. Right: Sketch of the three main exchange interactions (b) Photograph and (c) X-ray Laue image of the single crystal of SrMnGe$_2$O$_6$ grown by floating zone method.*

## II Experimental details

Single crystals of SrMnGe$_2$O$_6$ were grown using the floating zone furnace technique. To prepare SrMnGe$_2$O$_6$ powders the starting materials of reagent-grade SrCO$_3$, MnO$_2$, and GeO$_2$ were thoroughly ground in an agate mortar and pressed into pellets. The pellets were placed in a platinum boat and heated in air to 1100° C at 100 °C/h, then held at 1100 °C for 10 days and cooled down to room temperature.

The powder was sealed in a rubber tube, evacuated, and compacted into a rod (typically 4 mm in diameter and 10 cm long) using a hydraulic press under an isostatic pressure of 1 GPa. After removal from the rubber tube, the rods were sintered at 1100 °C for 5 days in air. Single crystals of approximately 4 mm in diameter and up to 50 mm in length were grown from the polycrystalline feed rods in a floating zone furnace. Growths were carried out under air at room pressure. The crystal growth rate was maintained at 4 mm/hr. Samples with dimensions suitable for the particular measurements and with different crystallographic orientations were cut from the same crystal. Both x-ray and neutron single crystal Laue diffraction confirmed the good crystallinity of the as-grown single crystals.

Magnetization was measured using a SQUID detection magnetometer (MPMS-XL by Quantum Design). The dc magnetic susceptibility measurements were performed under both zero-field-cooled (ZFC) and field-cooled (FC) procedures over the temperature range between 2 K and 300 K in magnetic field of 0.1 T.

Heat capacity measurement was carried out using a relaxation technique on a Quantum Design Physical Property Measurement System (PPMS). A small single crystal was mounted on a sample platform with Apiezon N grease for better thermal contact. The heat capacity was recorded in the temperature range of 2–300 K without external field.

The temperature dependence of electric polarization of $SrMnGe_2O_6$ was obtained by the pyroelectric current method. The oriented single crystals were coated with silver epoxy on both parallel surfaces of the sample. A poling field, Ep, up to ± 500 kV/m was first applied in the paraelectric state at temperature 6–10 K prior to cooling the sample through the Néel temperature down to 2 K in order to obtain a single polar domain state. At 2 K, the poling field was removed. Then, the sample was heated at a constant rate of 3 K/min, and the pyroelectric current curves were recorded using a Keithley 6514 electrometer. Electric polarization curves were eventually obtained by the integration of the time dependence of the observed pyroelectric current.

Neutron diffraction experiments were carried out on the CEA-CRG D23 4-circle single-crystal diffractometer at the Institut Laue Langevin (Grenoble, France) with an incident wavelength λ=1.277 Å selected by a fixed curvature Cu 200 monochromator. The single crystal was mounted in a close-cycle refrigerator and data collections were carried out at several temperatures in the paramagnetic domain and for each of the magnetic phases. The refinements were done using the FULLPROF SUITE package[23] by least-squares refinements using the integrated intensities and including an extinction correction following the model of Becker-Coppens [24]. To refine the magnetic structures, the crystallographic parameters and the scale factors were fixed to the values obtained in the crystalline refinements. The spin configurations have been described as follows: for a given magnetic propagation vector **k** (and the associated −**k**)

$$\mathbf{m}_{lj} = m_{1j}\hat{u}_j \cos[2\pi(\mathbf{k}.\mathbf{R}_l + \Phi_j)] + m_{2j}\hat{v}_j \sin[2\pi(\mathbf{k}.\mathbf{R}_l + \Phi_j)]$$

*Equation 1*

Where $\mathbf{m}_{lj}$ is the magnetic moment of the atom j in the unit cell l, $\mathbf{R}_l$ is the vector joining the arbitrary origin to the origin of unit cell l, and $\Phi_j$ is a magnetic phase. Group-theoretical calculations were done using ISODISTORT [25] and Bilbao crystallographic server (magnetic symmetry and application[26]) software.

The inelastic neutron scattering (INS) experiment was carried out on the 4F2 cold neutron three-axis spectrometer at the Laboratoire Léon Brillouin (ORPHÉE Reactor, Saclay, France). The spectrometer was equipped with focusing Pyrolitic Graphite PG (002) monochromator and analyzer. A Be filter was implemented in the scattered beam to remove high order contaminations. The final energy $E_f$ was set to 4.9 meV or 3.48 meV ($K_f$ =1.55 and 1.3 Å$^{-1}$, yielding an energy resolution of about 0.22 and 0.1 meV respectively). The INS measurement was performed on a long single crystal (21*2*2 mm$^3$, 176mg). The sample was mounted on a standard orange cryostat, and aligned in the (0 K L) scattering plane.

## III Results

### A. Magnetic and multiferroic properties

Temperature dependence of the magnetic susceptibility curves under a magnetic field along the *a*, *b* and *c\** axes of SrMnGe$_2$O$_6$ are shown in Figure 2a. The anomalies in these curves indicate the presence of two successive magnetic transitions. At $T_{N1}$=4.36 K, the first magnetic transition is clearly visible by a drop in the magnetic susceptibility along the *a* direction while only a small kink is observed along the *b* and *c\** directions. The second magnetic phase transition is featured by a sudden drop in the susceptibility along the *b* axis at $T_{N2}$=4.05 K (Figure 2b). The two successive magnetic transitions can be further confirmed by heat capacity measurements. As shown in Figure 2b, the specific heat curve exhibits a sharp peak and a lambda-like anomaly at $T_{N1}$ =4.36(2) K and $T_{N1}$ = 4.05(2) K, respectively, coinciding with the two magnetic transitions. Only one magnetic phase transition was observed in our previous study on powder probably because the two transitions are very close in temperature.

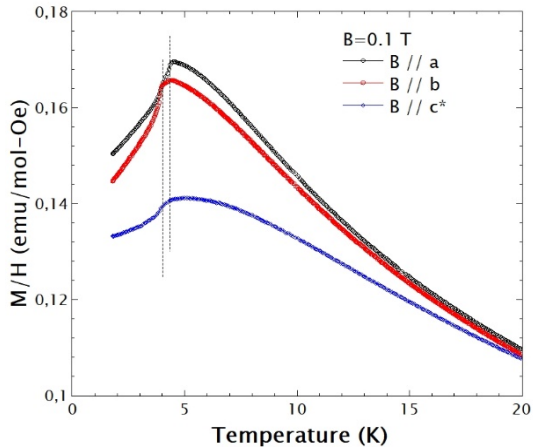

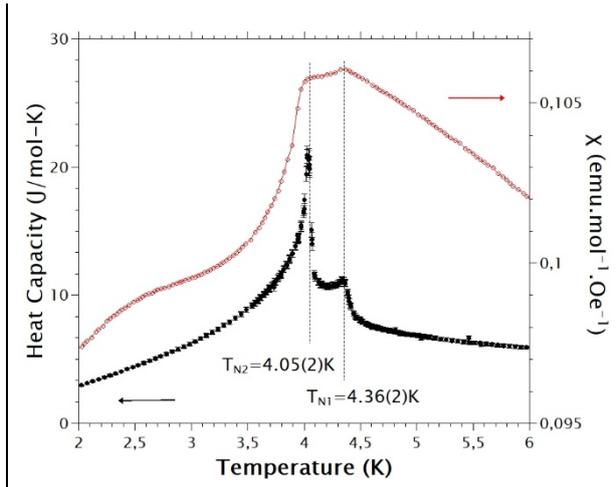

*Figure 2 (a) Temperature dependence of magnetic susceptibility measured under a magnetic field of 0.1 T applied along the a, b and c\* axis. (b) Temperature dependence of heat capacity and magnetic susceptibility measured under a magnetic field of 0.01 T applied along the b axis. The dashed lines indicate magnetic phase transitions.*

The electric polarization anisotropy measured on a single crystal often provides insights in the understanding of the pertinent mechanisms of multiferroicity. We have performed pyroelectric current measurements with the poling electric field applied along *a*, *b* and *c\** direction under $E_p = 500$ kV m$^{-1}$. As shown in Figure 3, the spontaneous electric polarization develops very close to 4 K and increases with decreasing temperature to reach ~3 μC m$^{-2}$ at 2 K along *a* and *c\** directions. On the other hand, there is minimal or no polarization seen along the *b* direction. The electric polarization value changes its sign with reversal of the electric poling direction which directly evidences that SrMnGe$_2$O$_6$ is indeed ferroelectric below 4 K. Moreover, the concomitant ferroelectric and antiferromagnetic transition at $T_{N2}$ = 4.05 K confirms that SrMnGe$_2$O$_6$ is a multiferroic compound, while the magnetic ordering at $T_{N1}$ = 4.36 K does not seem to cause a ferroelectric order.

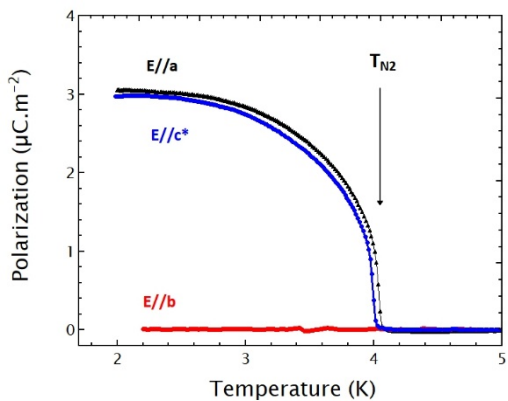

*Figure 3 Temperature dependence of the electric polarization of SrMnGe$_2$O$_6$ measured with the electric field parallel to the a, b and c\* axis.*

**B. Neutron diffraction**

The structural refinement based on single crystal neutron diffraction data confirmed that SrMnGe$_2$O$_6$ crystallizes with the C2/c symmetry, in good agreement with our previous results based on single crystal x-ray diffraction data[22]. No structural phase transition could be detected by our temperature-dependent neutron diffraction measurements down to 2 K. The atomic coordinates, displacement parameters and principal structural parameters are given in Table 1. Although the Mn-O1-Mn in-chain bond angle departs from 90° (Mn-O1-Mn = 96.12(6)°), the oxygen octahedra surrounding the Mn$^{2+}$ cations remain very regular as reflected by the small value of the distortion index based on bond lengths, D, defined by Baur[27].

*Table 1 Agreement factors and refined structural parameters for SrMnGe$_2$O$_6$ at 2K. The average distance, octahedral distortion and effective coordination number are also given for MnO$_6$ octahedron.*

| Name | x | y | z | B (Å$^2$) |
|---|---|---|---|---|
| Sr | 0 | 0.30855( 13) | 0.25 | 0.303( 28) |
| Mn | 0 | 0.90760( 23) | 0.25 | 0.072( 42) |
| Ge | 0.28244(6) | 0.09546( 8) | 0.21927( 13) | 0.169( 24) |
| O1 | 0.11107(9) | 0.08742( 11) | 0.13957( 19) | 0.344( 28) |
| O2 | 0.35783(10) | 0.25469( 11) | 0.31808( 21) | 0.394( 25) |
| O3 | 0.35351(8) | 0.02179( 11) | 0.98119( 20) | 0.312( 28) |

Mn-O1-Mn = 96.12(6)°

Mn-Mn distance through J: d= 3.257(3) Å

Mn-Mn distance through J1 : d$_1$= 5.996(2) Å

Mn-Mn distance through J2: d$_2$= 6.9896(19) Å

MnO$_6$ octahedron:

Average bond length = 2.1784 Å

Distortion index (bond length) = 0.01072

Effective coordination number = 5.9725

Space group C2/c; a=10.3559Å, b=9.3903Å, c=5.5133Å, β=104.651°, R$_{F2}$= 4.59, R$_{F2w}$=6.26, R$_F$=2.62, χ$^2$=118. Number of reflections: 582.

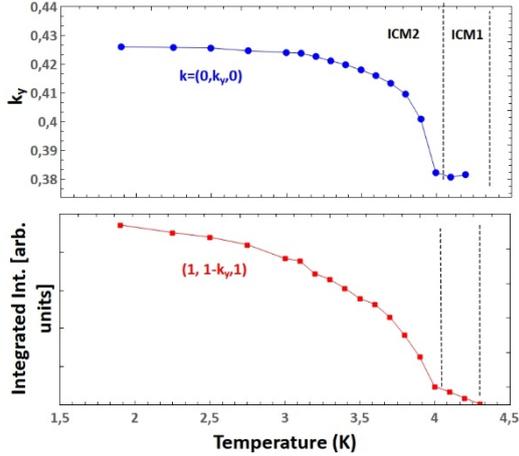

*Figure 4 Temperature dependence of the $k_y$ component of the magnetic propagation vector $k=(0,k_y,0)$ (top) and the integrated intensity of the (1, 1-$k_y$, 1) magnetic reflection (bottom).*

Single crystal neutron diffraction experiments show that magnetic Bragg reflections appear below $T_{N1}$ (ICM1 phase) which can be indexed by an incommensurate propagation vector **k**= (0, $k_y$, 0) with $k_y$=0.381(1) at 4.1 K. In Figure 4, one can clearly see that the modulus of the k-vector increases significantly below $T_{N2}$ to reach $k_y$=0.425(1) at 2 K, which marks the transition to the lower temperature ICM2 phase. This transition is also reflected by a kink in the evolution of the integrated intensity of (1, 1-$k_y$, 1) reflection. The k-vector corresponding to the ICM2 phase is consistent with the one previously observed by neutron powder diffraction[22]. To determine the magnetic structures, 248 reflections belonging to the ICM1 phase were collected at 4 K and 290 reflections were collected at 2 K for the ICM2 phase. Possible magnetic models where searched using the Simulated Annealing method and the corresponding magnetic structures were then refined. The two orbits Mn1 (0, 0.90760, 0.25) and Mn2 (0, 0.09240, 0.75) were constrained to have the same magnetic moment. Only the magnetic phase difference ΔΦ between these two sites was refined. The parameters that describe the proposed spin arrangements are gathered in Table 2.

The magnetic structure in the ICM1 phase was determined as an amplitude-modulated collinear sinusoidal structure with the reliability factor $R_f$ = 11%. The refinement leads to a magnetic moment of 1.961(6) $\mu_B$ along the easy-magnetic axis (Figure 5a). As shown in Figure 5c, the easy-magnetic axis is confined into the (*ac*) plane but with an angle of α =18.5° away from the *a* axis towards the c axis. The corresponding magnetic superspace group of the ICM1 phase is B2/b1'(0,0,g)s0s with g = $k_y$ ,[25,28] generated by the single magnetic irreducible representation mLD2 (see *Table 3*). Such magnetic symmetry preserves the two-fold axis and the mirror plane symmetry, resulting in the centrosymmetric magnetic point group 2/m1'. This explains the absence of spontaneous electric polarization between $T_{N1}$ and $T_{N2}$.

The determination of the magnetic structure of the ICM2 phase (reliability factor $R_f$ = 3.54%) leads to an elliptical cycloidal spin structure with the moments rotating within the plane formed by the *b* axis and the same easy-magnetic axis than in ICM1 phase (cf Figure 5a and c). It is worth noticing that the moments are close to be perpendicular to the *c* direction (4° off). Tentative refinements constraining them to be exactly normal to the *c* direction lead however to a much lower fit quality. At 2 K, the refined $Mn^{2+}$ magnetic moment varies from 4.26(2) $\mu_B$ along the long axis of the ellipse to 3.90(2) $\mu_B$ along the short axis. The magnetic superspace group describing this magnetic structure is Bb1'(0,0,g)0s where two magnetic irreducible representations mLD1 and mLD2 have to be combined to generate the magnetic structure. It breaks the structural two-fold axis and keeps the mirror plane symmetry perpendicular to the unique *b* axis, leading to the polar magnetic point group m1'. The corresponding magnetic point group allows the existence of spontaneous polarization in any direction perpendicular to the mirror, i.e. in the (*ac*) plane, in good agreement with the observed spontaneous polarization along *a* and *c\** (Figure 3).

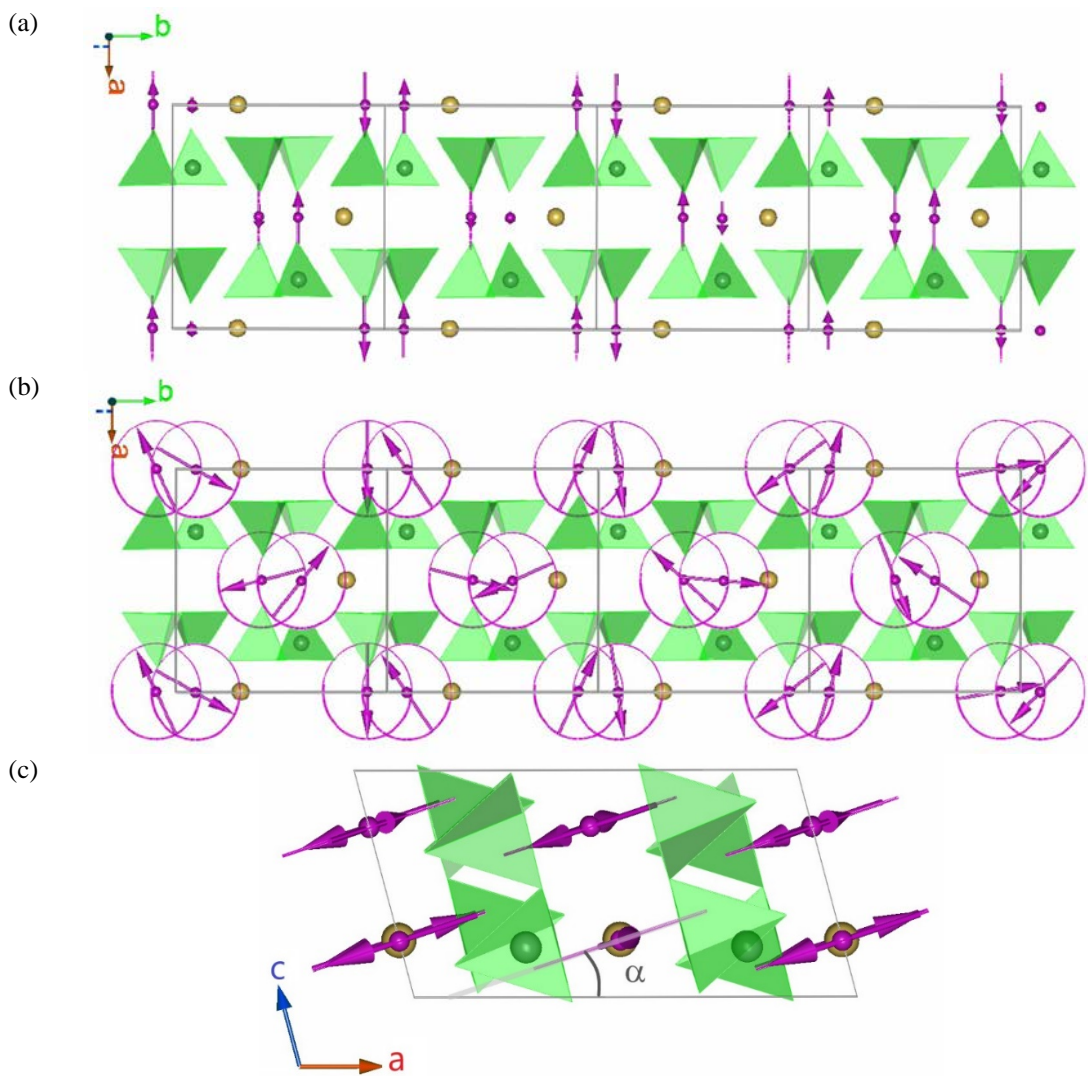

*Figure 5 : Illustration of the determined magnetic structures: (a) ICM1 amplitude-modulated collinear sinusoidal structure with moments along the direction defined by the α angle (b) ICM2 elliptical cycloidal spin structure with the moments rotating within the ab plane (c) α angle give the direction of the easy-magnetic axis located at about 18.5° from the a axis towards the c axis*

*Table 2 : Refined parameters of $SrMnGe_2O_6$ magnetic structures at 2 K and 4.1K under zero field. Both structure have a propagation vector of the form k= [0, $k_y$,0]. The unitary vectors $\hat{u}$ and $\hat{v}$ are described by spherical angles: φ is the angle that the projections of the unitary vectors in the xy plane make with x(//a) and θ is the angle the unitary vectors make with z (//c∗). ΔΦ corresponds to the magnetic phase difference of Mn1 (0, 0.90760, 0.25) and Mn2(0, 0.09240, 0.75). α is the angle between the plane where the moments lay and a direction in the ac plane. $m_{1j}$ and $m_{2j}$ correspond to the modulus of the orthogonal components of magnetic moments. $R_{F2}$, $R_{F2w}$, $R_F$, and $χ2$ are the reliability factors.*

| Temperature | | 2K | 4.1K |
|---|---|---|---|
| Magnetic Superspace group | | Bb1'(0,0,g)0s | *B2/b1'*(0,0,g)s0s |
| Basis | | {(-1,0,0,0),(0,0,-1,0),(0,-1,0,0),(0,0,0,1)} | |
| Propagation vector | | [0,$k_y$,0] | |
| $k_y$ | | 0.425 | 0.381 |
| Mn1, Mn2 | $m_{1j}$ (μ$_B$) | 4.26(2) | 1.961(6) |
| | $φ_u$ | 0° | 0 |
| | $θ_u$ | 71.4(3)° | 71.5(4) |
| | $m_{2j}$ (μ$_B$) | 3.90( 2) | |
| | $φ_v$ | 90° | |
| | $θ_v$ | 90° | |
| ΔΦ | | 0.1744( 6) | 0.2089( 15) |
| α | | 18.6(3)° | 18.5(4)° |
| $R_{F2}$ (%) | | 4.37 | 12.5 |
| $R_{F2w}$ (%) | | 4.79 | 10.2 |
| $R_F$ (%) | | 3.54 | 11.0 |
| $χ^2$ | | 3.79 | 2.73 |
| Number of Reflections | | 290 | 248 |

*Table 3 : Nonzero IR's and associated basis vectors ψ for the space group C2/c with k = [0 0.424 0]. The magnetic atoms $Mn^{2+}$ at 4e site are split into two independent orbits: Mn1 (0, 0.90760, 0.25) and Mn2 (0, 0.09240, 0.75).*

| | | Orbit 1 | | | Orbit 2 | | |
|---|---|---|---|---|---|---|---|
| IR | Basis vector | $m_x$ | $m_y$ | $m_z$ | $m_x$ | $m_y$ | $m_z$ |
| mLD1 | ψ1 | 0 | 1 | 0 | 0 | 1 | 0 |

| | | | | | | | | |
|---|---|---|---|---|---|---|---|---|
| mLD2 | ψ2 | 1 | 0 | 0 | 1 | 0 | 0 | |
| | ψ3 | 0 | 0 | 1 | 0 | 0 | 1 | |

## C. Magnetic phase diagram

As explained above the dominant magnetic interaction (J) between the $Mn^{2+}$ nearest-neighbor cations (d=3.257Å) along the octahedra chains is mediated by super-exchange interactions through the $O_1$ oxygen anion in competition with direct exchange interactions. The magnetic coupling between $MnO_6$ chains is operated through the bridging $GeO_4$ tetrahedra (see Figure 6). Two magnetic interactions $J_1$ and $J_2$ are expected to play an important role for diverse magnetic orderings: $J_1$ is governed by a double super-super-exchange (SSE) path through edges of two different $GeO_4$ tetrahedra ($d_1$=5.996Å), while $J_2$ is given by a single SSE at a longer distance ($d_2$=6.989Å).

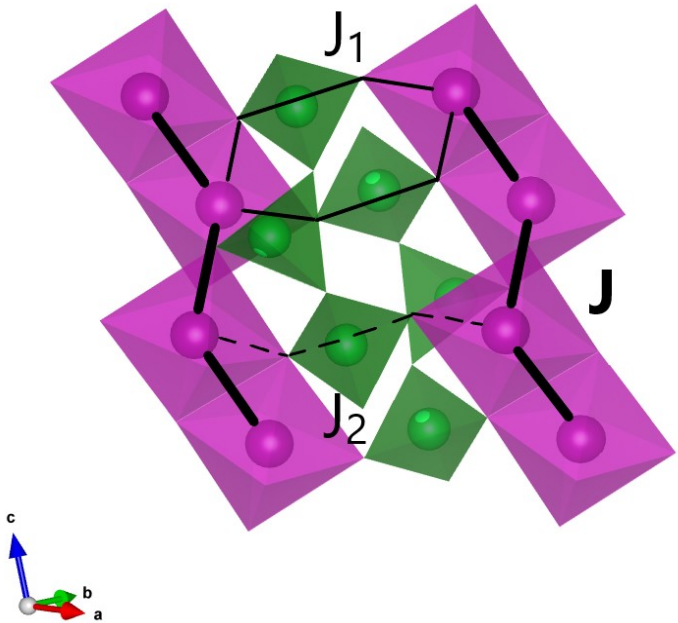

*Figure 6 : The $MnO_6$ zigzag chains connected through $GeO_4$ tetrahedra in $SrMnGe_2O_6$. The bold solid, thin solid and dotted lines correspond to exchange paths J, $J_1$ and $J_2$.*

This model with the three isotropic[29] exchange interactions (J, J1, J2) was solved assuming classical spins to determine the ordered spin configurations in $SrMnGe_2O_6$. The Luttinger-Tisza method[30] was used for finding the ordering wave-vector $q_0$ that minimizes the total classical energy $E(\boldsymbol{q}) = -\mathcal{N}|\boldsymbol{S}(\boldsymbol{q})|^2 \lambda_0(\boldsymbol{q})$ (with $\mathcal{N}$ the number of spins and $S(q)$ their Fourier components), ie that maximizes the highest eigenvalue $\lambda_0(\boldsymbol{q})$ of the Fourier transform of the interaction matrix $\mathbf{J}(\mathbf{q})$, for given sets of exchange couplings J, $J_1$, $J_2$. The resulting classical phase diagram in the ($J_1$/|J|, $J_2$/|J|) is represented in Figure 7. When one of the interchain interaction dominates the other, commensurate orders are found: $J_1$ promotes **k**=0 while $J_2$ tends to break the centering C with a **k**=(0, 1, 0) vector. The **k**=0 commensurate order is found in most of the pyroxenes with the same crystal structure of space group *C2/c* with AFM J: in $NaCrGe_2O_6$[31], $NaCrSi_2O_6$ [32] and $CaMnGe_2O_6$[20]. Incommensurate order comes out from the competition between $J_1$ and $J_2$ interchain interactions, it represents a large intermediate sector of the phase diagram. The stabilized incommensurate k vector is (0, $k_y$,0). It is worth noticing that in some part of the diagram ($k_x$, 0, $k_z$) order is very close in energy and may be further stabilized by introducing a small antiferromagnetic next nearest-neighbor (NNN) interaction in the chain. This is probably the case with NaFeGe$_2$O$_6$ [14]. Note that **k**=(0, 1, 0) is found in all pyroxenes where the intrachain interaction is ferromagnetic: CaM(Si,Ge)$_2$O$_6$ with M= Fe, Co, Ni[33] and SrCoGe$_2$O$_6$[22].

To get more insight into the phase diagram, we have also calculated analytically the classical energy for the **k**=(0, $k_y$, 0) propagation vector. Minimizing the classical energy with respect to $k_y$ leads to the following relation between $k_y$ and the exchange interactions $J_1$ and $J_2$:

*Equation 2*

$$J_2 = \frac{J_1}{2\sqrt{J^2 + 2J_1 \cos(k_y/2) + J_1^2}}$$

for $J_1$ and $J_2$ <0 (AFM). The line displayed on Figure 7 illustrates this relationship for $k_y$=0.424. Taking into account the phase ($\Delta\Phi$) between the magnetic moment directions of the two magnetic sites gives another parameterization:

$$\frac{J_1}{|J|} = -\frac{\sin(k_y + \Delta\Phi)}{\sin(\frac{k_y}{2} + \Delta\Phi)}$$

$$\frac{J_2}{|J|} = -\frac{\sin(k_y + \Delta\Phi)}{2\sin(\frac{k_y}{2})}$$

If we now consider the actual structure refined in SrMnGe$_2$O$_6$ at 2 K ($k_y$=0.424(1), $\Delta\Phi$ = 62.8(2)° ), it gives $J_1/|J|$=0.88(2) and $J_2/|J|$=0.29(1) which is represented by the black point marked on Figure 7.

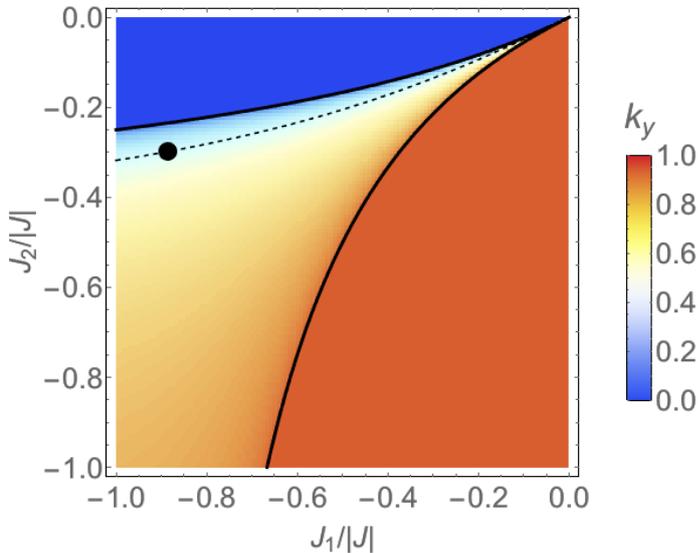

*Figure 7 : Magnetic phase diagram for an antiferromagnetic intrachain interaction J<0, representing the stability of different magnetic ground states in function of on relative values of the exchange parameters, $J_1/|J|$ and $J_2/|J|$. Phase boundaries are represented by the plain black lines while the dashed one shows the iso-$k_y$=0.424. The black point is the sets of parameters obtained for SrMnGe$_2$O$_6$ (see text). The color scale represents the $k_y$ value: the red region shows the magnetic phase characterized by k = (0, 1, 0), while the region in blue corresponds to the k = 0 phase.*

## D. Spin-wave excitations from inelastic neutron scattering

To cross check this analysis, we have carried out inelastic neutron scattering (INS) experiment at 2 K. Representative raw data scans are shown in Figure 8.

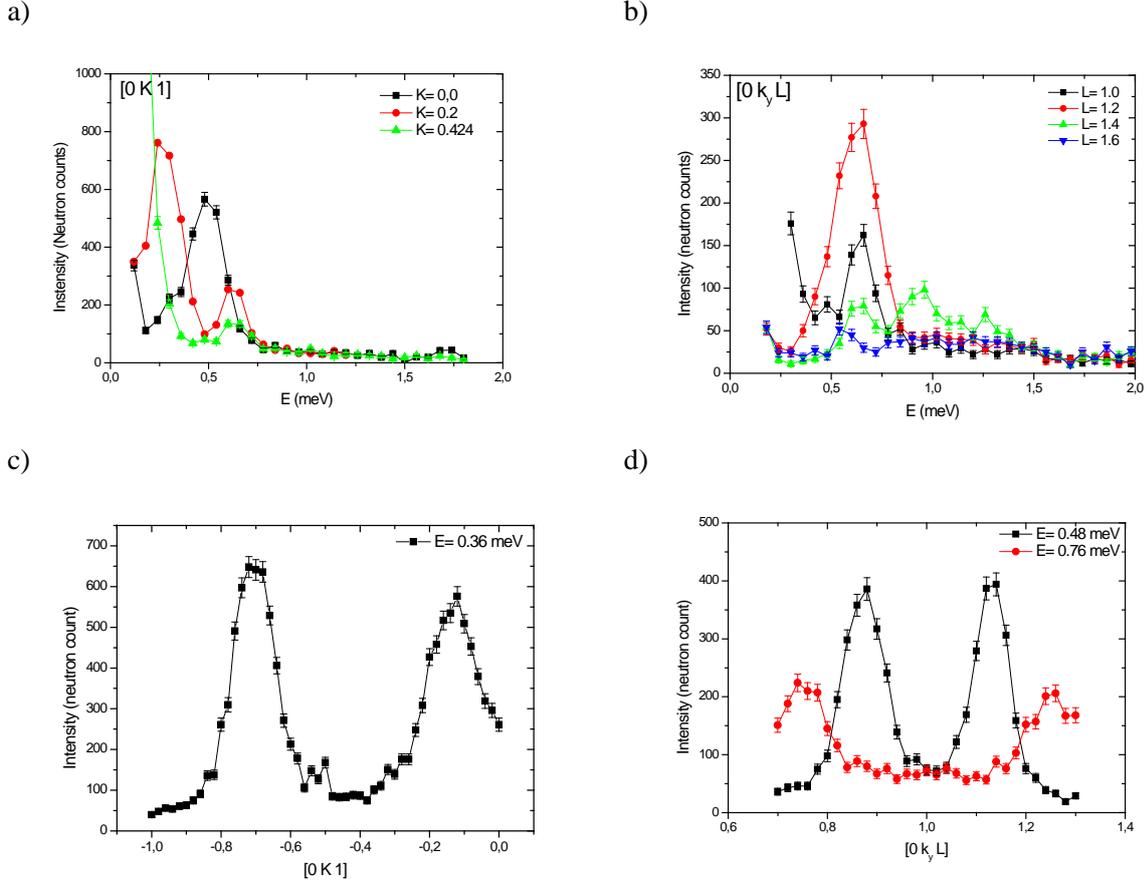

*Figure 8 : Energy scans measured at selected positions Q=(0 $k_y$ L) (a) and Q=(0 K 1)(b) to determine the magnon dispersion in the ferroelectric elliptical cycloidal phase at T= 1.5 K. (c) and (d) Representative Q-scans at constant energy transfer in the [0 K 1] (c) and [0 ky L] (d) directions crossing the spin-waves excitations.*

Well-defined spin-wave excitations are observed, as shown also in Figure 9. Acoustic modes emerge from the incommensurate (0,$k_y$,1) Bragg peaks and disperse throughout the Brillouin zone up to about 1 meV, as expected in such a non-collinear structure. Along L, a more complicated spectrum emerges: the acoustic modes go soft at integer L=1 and 2, while an optical mode is also observed at about 0.7 meV.

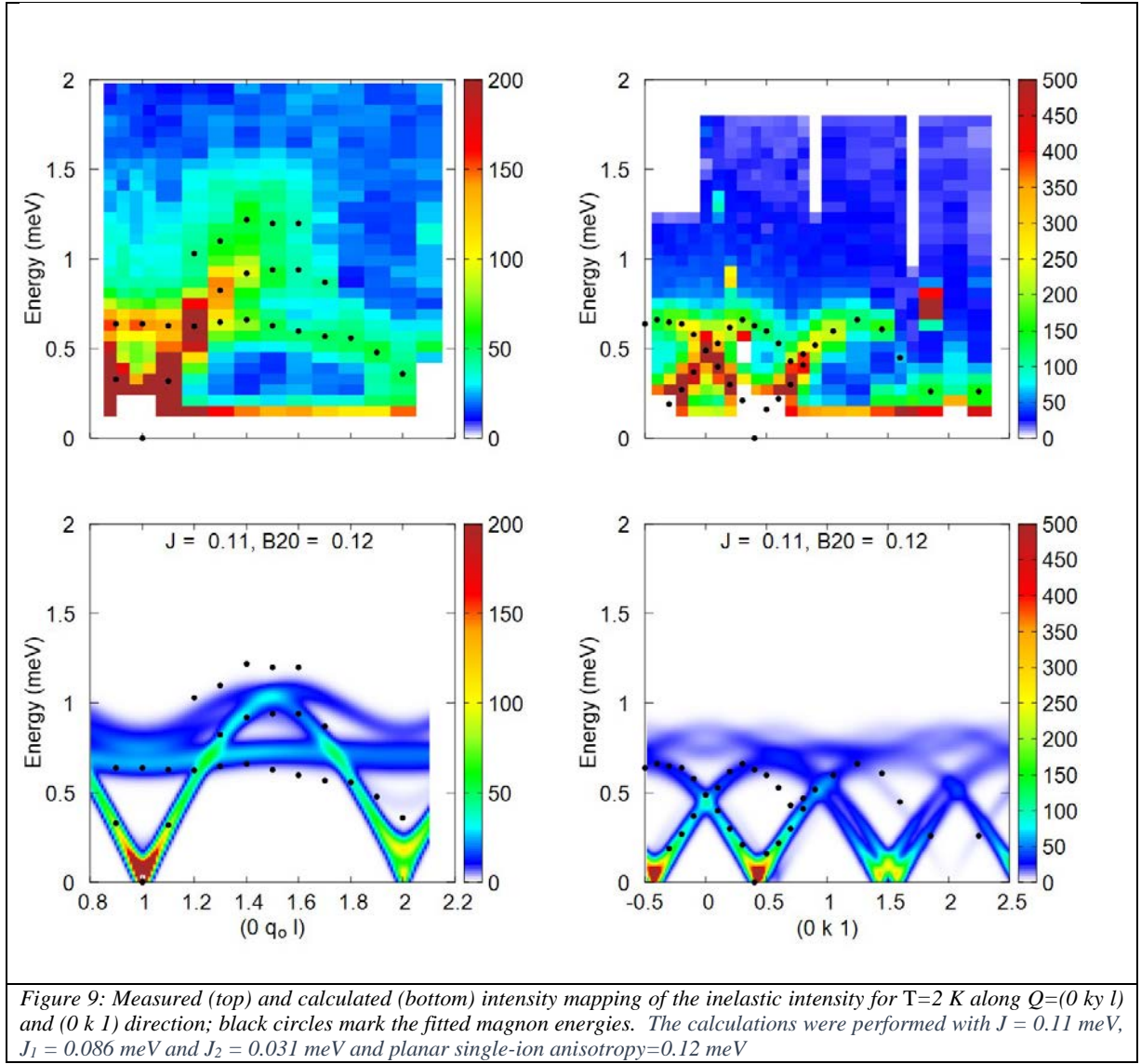

*Figure 9: Measured (top) and calculated (bottom) intensity mapping of the inelastic intensity for T=2 K along Q=(0 ky l) and (0 k 1) direction; black circles mark the fitted magnon energies. The calculations were performed with J = 0.11 meV, $J_1$ = 0.086 meV and $J_2$ = 0.031 meV and planar single-ion anisotropy=0.12 meV*

To model the spin dynamics, spin-wave calculations were performed using the *Spinwave* software[34] developed at LLB. The Hamiltonian governing the magnetic properties is of the form:

$$H = J \sum_{i,j}^{in\ chain} S_i S_j + J_1 \sum_{i,j}^{inter\ chain} S_i S_j + J_2 \sum_{i,j}^{inter\ chain} S_i S_j + \sum_i B_{i,20} \vartheta_{i,20}$$

In addition, to the J, $J_1$ and $J_2$ terms, a single-ion anisotropy term was taken into account, modeled by V = 3/2 $B_{20}$ $S_{z,i}^2$. Here $B_{20}$ is positive to ensure that the spins lie in the easy plane and stabilize the cycloid structure, as inferred from diffraction. Diagonalization of the Hamiltonian in the spin wave approximation allows to calculate the spin-spin correlation function as observed by inelastic neutron-scattering experiments. As we have seen before, considering the actual structure refined in $SrMnGe_2O_6$ gives constraints on the ratios $J_1/|J|$ and $J_2/|J|$ to fulfill Equation 2. The values of J and $B_{20}$ have yet to be determined. To this end, we have carried out a series of calculations keeping the ratios $J_1/|J|$ and $J_2/|J|$ constant, but varying J and $B_{20}$ in a systematic way. The detailed study is presented in Appendix A. The best reasonable agreement is found for the following set of interactions consistent with the refined magnetic structure i.e. J = 0.11(1) meV, J1 = 0.086(12) meV and J2 = 0.031(4) meV and $B_{20}$=0.12(3) meV, see Figure 9, hence showing the consistency of this analysis against our neutron results.

**IV. Discussion**

We have just seen that the refined magnetic structure, the calculated magnetic phase diagram and the spin-wave measured in $SrMnGe_2O_6$ are well captured by the simple model of three isotropic exchange interaction J, $J_1$ and $J_2$. To the best of our knowledge, this study shows the first experimental determination of exchange interactions done on pyroxene compounds. DFT calculations were however performed on isostructural compounds with $3d^5$ magnetic cation: $CaMnGe_2O_6$[35] and $NaFeGe_2O_6$[14]. The hierarchy and the sign of the exchange interactions is the same in the three compounds: all interactions are antiferromagnetic and the strongest is the exchange interaction along the chains (Super-exchange J), while the second largest exchange coupling is via the two $GeO_4$ tetrahedra (Super-super-exchange SSE J1). This hierarchy can be explained by (i) the nature of the interaction SE vs SSE, (ii) the Mn-Mn distances and (iii) the fact that J1 involves two $GeO_4$ bridge compared to only one for J2. $SrMnGe_2O_6$, has much weaker exchange interaction values than in the other two compounds ($J_{NaFeGe2O6}$=12.3K, $J_{CaMnGe2O6}$=3.6K, $J_{SrMnGe2O6}$ =1.04K), which is reflected also in the Néel ordering temperatures ($T_{N\ NaFeGe2O6}$=13K, $T_{N\ CaMnGe2O6}$= 15K, $T_{N\ SrMnGe2O6}$ =4.36K). The weaker exchange interaction in $SrMnGe_2O_6$ can be understood by considering the distortion of the ideal pyroxene structure due to the large $Sr^{2+}$ cation size. First of all, in $SrMnGe_2O_6$ the magnetic cation distances M-M in between the chains (d1 through J1 and d2 through J2) are much larger than in $NaFeGe_2O_6$ (for instance $d_1$= 5.996 Å in $SrMnGe_2O_6$ with respect to 5.64Å in $NaFeGe_2O_6$[9]). Then, let us consider the strength of the dominant intrachain magnetic interaction J in $SrMnGe_2O_6$. Super-exchange through the O1 oxygen anion between the $Mn^{2+}$ nearest-neighbor cations along the octahedra chains is in competition with direct exchange. The detailed analysis of the different orbital contribution done by Streltsov and Khomskii[21] show that $t_{2g}$–$e_g$ contribution is the strongest and is AFM. $e_g$–$e_g$ orbitals give a smaller contribution but which is subtly dependent on the angle M-O-M. For angle close to 90° FM contribution dominates, while it is outbalanced by AFM for much larger angle, the compensation occurring about 97°. In $SrMnGe_2O_6$, the angle M–O1–M is 96.12(6)°, close to the compensation angle, and therefore $e_g$-$e_g$ orbital have low or no contribution. In $NaFeGe_2O_6$ however, the M-O-M angle is ~102.7° and the $e_g$-$e_g$ contribution reinforces the AFM character of the interaction. This is an important difference between the compounds because while for $NaFeGe_2O_6$[14] the value of intra-chain J is nearly three times larger than inter-chain interactions J, in $SrMnGe_2O_6$ J and $J_1$ are comparable ($J_1$/J =0.88). This raises the question of whether $NaFeGe_2O_6$ is in the quasi-one-dimensional limit unlike $SrMnGe_2O_6$. This is a non-trivial issue because, the connectivity is quite peculiar in pyroxenes. While J connects the spins of a given chain running along the c-axis, $J_1$ connects two spins alternately within the neighboring left, right, front and back chains. In this sense, even with J and $J_1$ comparable, SrMnGe$_2$O$_6$ is not fully 3D.

Several experimental facts evidence that magnetic anisotropy is an important feature in SrMnGe$_2$O$_6$. Firstly, the magnetization susceptibility measured along the *c\** direction is much smaller than along *b* and *a*, which is consistent with *c\** being a hard axis of magnetization. Then, the presence of anisotropy promotes a collinear structure over a non-collinear one which is the case here where at high temperature an amplitude-modulated collinear sinusoidal structure is preferred. Furthermore, the magnetic structures refined in both phases present an easy-axis direction in the (*a,c*) plane, almost perpendicular to *c\**, with a robust direction with respect to temperature variations. Planar single-ion anisotropy was introduced in the spin-wave calculation to stabilize the cycloid structure in the observed plane. The observed anisotropy is significant since in a purely ionic description manganese in SrMnGe$_2$O$_6$ is an S-state Mn$^{2+}$ ion with vanishing orbital moment. However it has been reported that single-ion anisotropy of Mn$^{2+}$ cannot be neglected and play an important role[36].

Having established the microscopic magnetism model and the driving factors stabilizing the polar elliptical cycloidal magnetic phase in SrMnGe$_2$O$_6$, let us now discuss in detail the direction of the electric polarization in the (ac) plane and the mechanism that can be proposed to generate it. We have measured a polarization along the *a* and *c\** directions with almost the same amplitude, which gives a resultant experimental polarization P$_{exp}$ with an angle of ~45 ° with respect to the *a* direction in the (*ac*) plane (see Figure 10). In the well-known spin current[1] and inverse DM theories[2,37], for two adjacent spins S$_i$ and S$_j$ separated by the vector e$_{ij}$ the polarization is expressed by $P_1 \propto e_{ij} \times (S_i \times S_j)$, and is therefore lying along the direction given by α. These mechanisms cannot therefore explain fully the observed polarization. We have to invoke also a polarization parallel to the cross-product $P_2 \propto S_i \times S_j$, via the extended inverse DM effect proposed by Kaplan and Mahanti[6]. It is indeed allowed by symmetry because there is neither a mirror plane containing e$_{ij}$ nor an n-fold rotation axis perpendicular to e$_{ij}$. Finally, in SrMnGe$_2$O$_6$, the induced macroscopic polarization can be understood in terms of the inverse Dzyaloshinskii-Moriya interaction combining the two orthogonal components P$_1$ and P$_2$. This combination was found also other spin-driven multiferroics like in delafossite AgFeO$_2$[38] or in RbFe(MoO$_4$)$_2$[39].

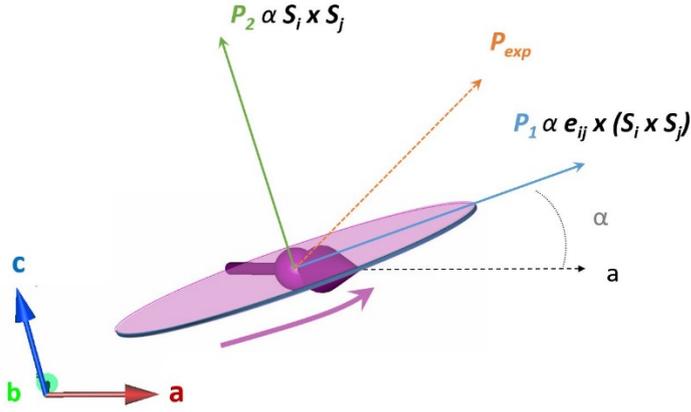

*Figure 10 : Schematic illustration representing the elliptical cycloid running along the b axis and the electric polarization directions in the (a,c) plane: $P_{exp}$ being the experimental one, $P_1$ and $P_2$ corresponding to the extended inverse DM mechanism[6].*

## V. Conclusion

In summary, we have performed a comprehensive study of multiferroic properties of $SrMnGe_2O_6$. We found two successive incommensurate spin structures below 4.36(2) and 4.05(2)K. A non-polar amplitude-modulated collinear sinusoidal magnetic structure emerges followed by a polar elliptical cycloidal spin structure. Extended Dzyaloshinsky-Moriya mechanism successfully explains the appearance of a spontaneous polarization measured within the (*ac*) plane associated with the latter magnetic symmetry. The good adequacy between the determined magnetic model, the calculated magnetic phase diagram and the measured and simulated spin-wave demonstrates that the magnetic behavior of $SrMnGe_2O_6$ is well captured by a rather simple model with three competing isotropic interactions and a weak single-ion anisotropy.

## APPENDIX A: Analysis of INS data

As explained in the main text, the numerical approach described in this study is based on the following Hamiltonian:

$$H = J \sum_{i,j}^{in\ chain} S_i S_j + J_1 \sum_{i,j}^{inter\ chain} S_i S_j + J_2 \sum_{i,j}^{inter\ chain} S_i S_j + \sum_i B_{i,20}\, \vartheta_{i,20}$$

In addition, to the J, $J_1$ and $J_2$ terms, a single-ion anisotropy is taken into account, modeled by a $B_{20} O_{20}$ term, with $O_{20} = 3/2\ S_z^2 - 5/2(5/2+1)Id$. This operator is written with respect to a local "z" quantification axis, along the (-0.33, 0, 1) vector. $B_{20}$ is positive to ensure that the spins lie in the plane perpendicular to this axis, and stabilize the cycloid structure, as inferred from diffraction. Meanwhile, the spin wave theory is a theory of harmonic deviations of the spins around a mean field solution. At the level of this approximation, the incommensurate propagation vector $k_y$ and the $\Delta\Phi$ phase shift are related via the formula:

$$\frac{J_1}{|J|} = -\frac{\sin(k_y + \Delta\Phi)}{\sin(\frac{k_y}{2} + \Delta\Phi)}$$

$$\frac{J_2}{|J|} = -\frac{\sin(k_y + \Delta\Phi)}{2\sin(\frac{k_y}{2})}$$

Diffraction measurements carried out in SrMnGe$_2$O$_6$ at 2 K yield $k_y$=0.424(1), $\Delta\Phi$ = 62.8(2)°, hence $J_1/|J|$=0.88(2) and $J_2/|J|$=0.29(1). Provided our minimal model is correct, these measurements thus put sever constraints on the uncertainties relative to $J_1/|J|$ and $J_2/|J|$. The values of J and B$_{20}$ have yet to be determined.

To this end, a series of calculations has been carried out keeping the ratios $J_1/|J|$ and $J_2/|J|$ constant, but varying J and B$_{20}$ in a systematic way. We have considered 0.05 < J < 0.19 meV and 0 < B$_{20}$ < 0.42 meV. At the same time, we compared the Curie-Weiss temperature value predicted by the calculation to the experimental value, equal to 20K. The main results are reproduced in Figures A1 to A4. These figures display the neutron intensity calculated along (0 q$_0$ 1) and (0 k 1), as well as the measured energies of the modes based on standard fits to the experimental data (see the black dots in the figures). The full experimental data are also reproduced for a more convenient comparison. These calculations show relatively good agreement for 0.1 < J < 0.12 meV with 0.06 < B$_{20}$ < 0.18 meV. More precisely, best values are found for J = 0.1 meV, B$_{20}$ = 0.15, and, if J = 0.12, we get B$_{20}$ = 0.09 meV. This indicates a correlation between J and B$_{20}$, estimated to be B$_{20}$ ~ -3(J-0.1)+0.15. According to this analysis, we thus have:

J = 0.11 +/- 0.01 meV and B$_{20}$ ~ -3(J-0.1)+0.15, i.e. and $\delta$B$_{20}$ ~ 3 $\delta$J.

Figure A4 displays part of the calculations performed for the (0 k 1) direction. They confirm the relatively good agreement for the range of parameters determined above.

Figure A5 shows the Curie Weiss temperature calculated in the (J, B$_{20}$) range of interest. The comparison with the experimental value is quite good, hence giving some more confidence in the analysis. Since the experimental Curie-Weiss temperature is 20K, this cross-check calculation would favor the strongest values of J, hence J=0.12 meV, B$_{20}$ = 0.09 meV.

We also should stress that in incommensurate magnets, in-plane and out-of-plane modes are expected. In-plane modes correspond to correlations between spin components within the spiral plane. In the long wavelength limit, they correspond to a global phase shift of the spins. These are the Goldstone mode of the model, and soften down to zero energy at the magnetic Bragg peaks. Importantly, the planar anisotropy term $B_{20}$ does not induce any spin gap for them since the spins keep an overall degree of rotation in the easy plane. In contrast, Figures A1 to A4 show that $B_{20}$ affects quite strongly the high-energy part of the spectrum.

We anticipate, however, that measurements in other directions are likely necessary to come to a definitive conclusion and propose a better model.

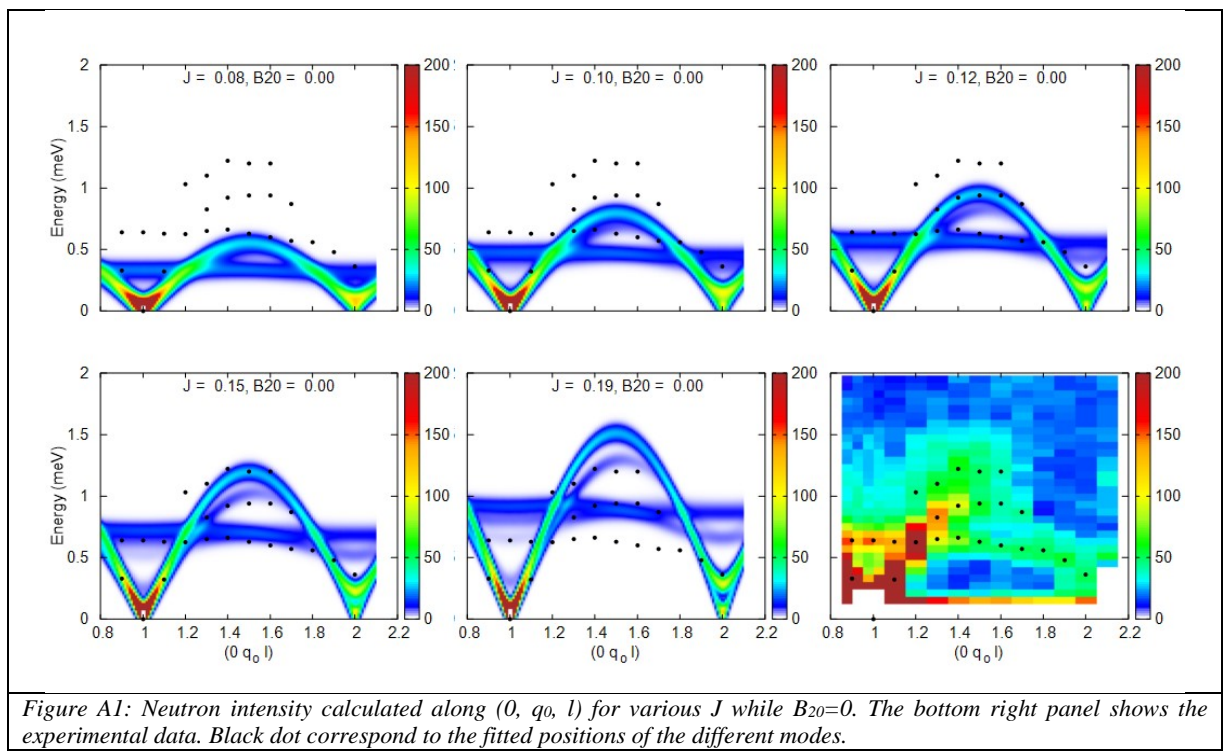

*Figure A1: Neutron intensity calculated along (0, $q_0$, l) for various J while $B_{20}$=0. The bottom right panel shows the experimental data. Black dot correspond to the fitted positions of the different modes.*

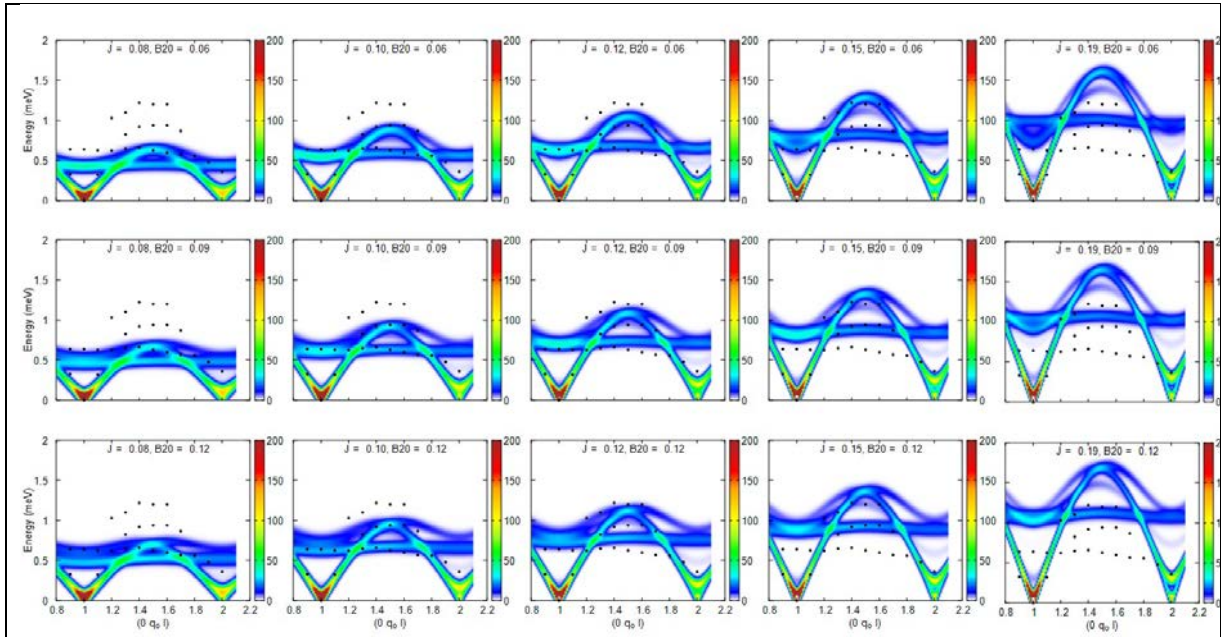
*Figure A2: Neutron intensity calculated along (0, $q_0$, l) for various J and $B_{20}$.*

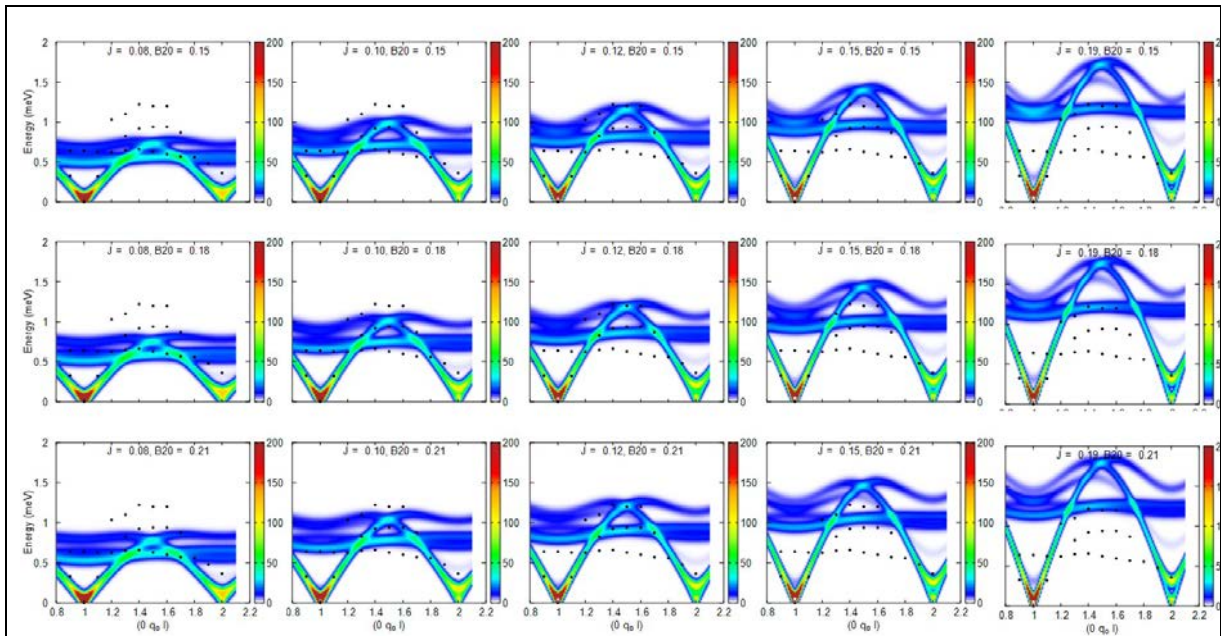
*Figure A3: Neutron intensity calculated along (0, $q_0$, l) for various J and $B_{20}$.*

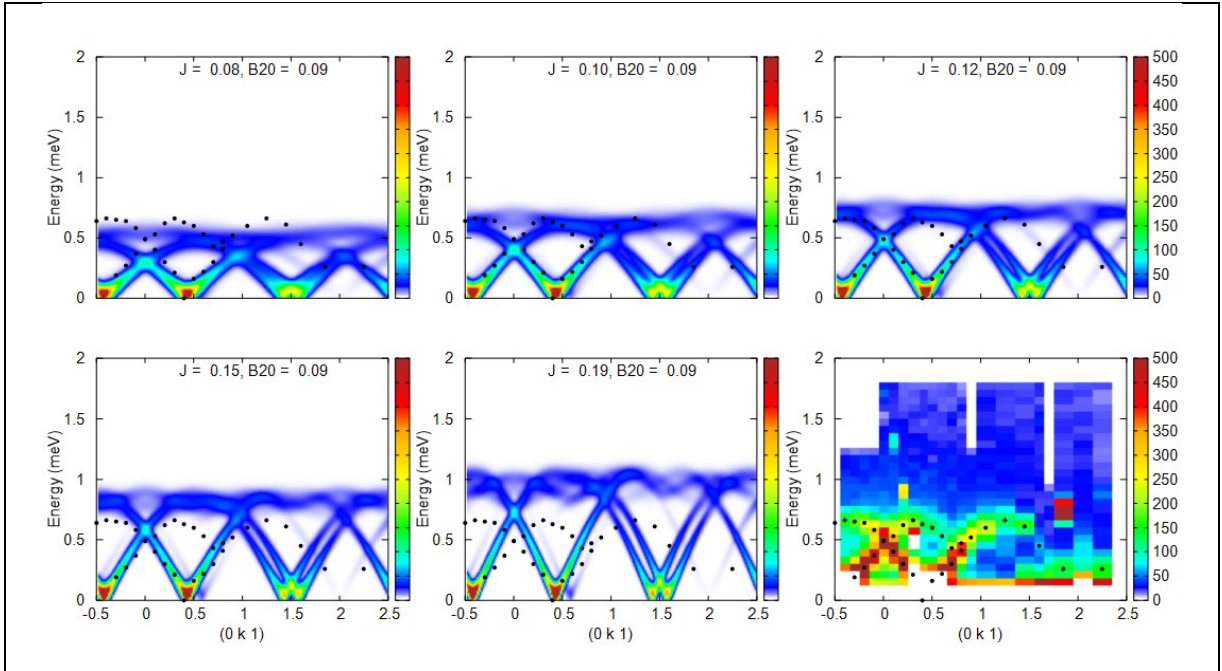

*Figure A4: Neutron intensity calculated along (0, k, 1) for various J and $B_{20}$ = 0.09 meV. The bottom right panel shows the experimental data.*

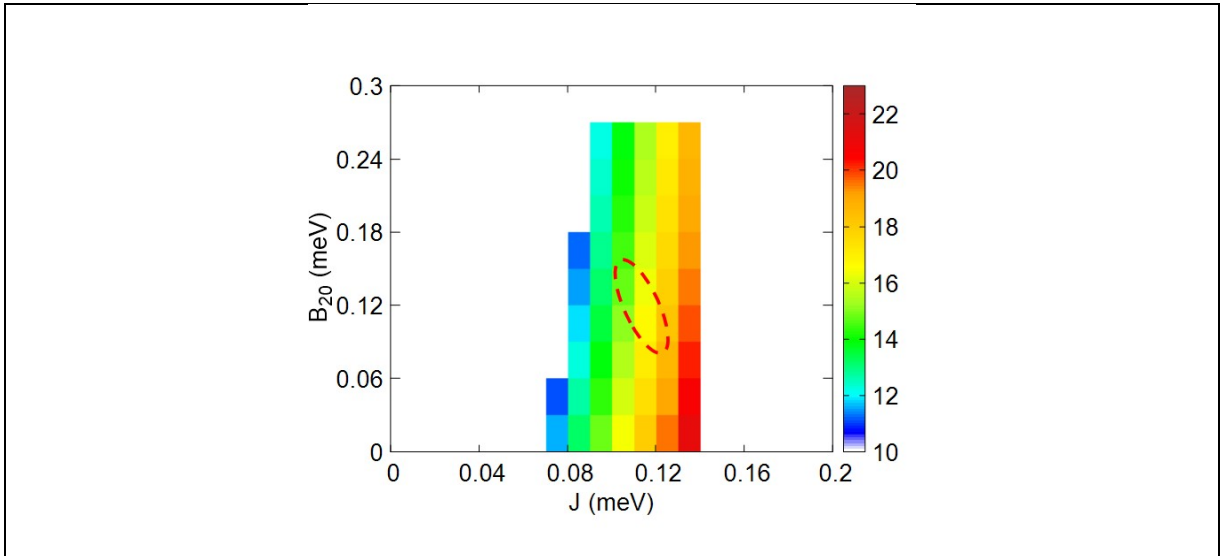

*Figure A5: Curie Weiss temperature (in K, see the color scale on the right of the figure). The dashed ellipse shows the range of parameters determined by comparison with INS data. Since the experimental Curie-Weiss temperature is 20K, this cross-check calculation favors the strongest values of J.*

**Acknowledgement:**

This work was financially supported by the ANR-13-BS04-0013. We acknowledge J. Debray, A. Hadj-Azzem and J. Balay for their help in the preparation of the samples and Andres Cano for fruitful discussions. We gratefully acknowledge the Fédération Française de la Neutronique (2FDN) and Institut Laue Langevin (ILL) (Grenoble, France) for beam time access to the fabulous D23. L.D. acknowledges support from the Rutherford International Fellowship Programme (RIFP). This project has received funding from the European Union's Horizon 2020 research and innovation program under Marie

Skłodowska- Curie Grant Agreement No. 665593 awarded to the Science and Technology Facilities Council.

[1] H. Katsura, N. Nagaosa, and A. V. Balatsky, Phys. Rev. Lett. **95**, 057205 (2005).

[2] I.A. Sergienko and E. Dagotto, Phys. Rev. B - Condens. Matter Mater. Phys. **73**, 094434 (2006).

[3] T.H. Arima, J. Phys. Soc. Japan **76**, 073702 (2007).

[4] M. Kenzelmann, A.B. Harris, S. Jonas, C. Broholm, J. Schefer, S.B. Kim, C.L. Zhang, S.W. Cheong, O.P. Vajk, and J.W. Lynn, Phys. Rev. Lett. **95**, 087206 (2005).

[5] Y. Yamasaki, S. Miyasaka, Y. Kaneko, J.-P. He, T. Arima, and Y. Tokura, Phys. Rev. Lett. **96**, 207204 (2006).

[6] T.A. Kaplan and S.D. Mahanti, Phys. Rev. B - Condens. Matter Mater. Phys. **83**, 174432 (2011).

[7] S. Jodlauk, P. Becker, J.A. Mydosh, D.I. Khomskii, T. Lorenz, S. V. Streltsov, D.C. Hezel, L. Bohat́, L. Bohatý, L. Bohaty, and L. Bohat́, J. Phys. Condens. Matter **19**, 432201 (2007).

[8] S. V. Streltsov, J. McLeod, A. Moewes, G.J. Redhammer, and E.Z. Kurmaev, Phys. Rev. B **81**, 045118 (2010).

[9] G.J. Redhammer, A. Senyshyn, M. Meven, G. Roth, S. Prinz, A. Pachler, G. Tippelt, C. Pietzonka, W. Treutmann, M. Hoelzel, B. Pedersen, and G. Amthauer, Phys. Chem. Miner. **38**, 139 (2011).

[10] M. Baum, A.C. Komarek, S. Holbein, M.T. Fernández-Díaz, G. André, A. Hiess, Y. Sidis, P. Steffens, P. Becker, L. Bohatý, and M. Braden, Phys. Rev. B **91**, 214415 (2015).

[11] I. Kim, B.-G. Jeon, D. Patil, S. Patil, G. Nénert, and K.H. Kim, J. Phys. Condens. Matter **24**, 306001 (2012).

[12] M. Ackermann, L. Andersen, T. Lorenz, L. Bohatý, and P. Becker, New J. Phys. **17**, 13045 (2015).

[13] T. Drokina, G. Petrakovskii, L. Keller, and J. Schefer, J. Phys. Conf. Ser. **251**, 012016 (2010).

[14] L. Ding, P. Manuel, D.D. Khalyavin, F. Orlandi, and A.A. Tsirlin, Phys. Rev. B - Condens. Matter Mater. Phys. **98**, 094416 (2018).

[15] C. Lee, J. Kang, and J. Hong, Chem. … **26**, 1745 (2014).

[16] P.J. Baker, H.J. Lewtas, S.J. Blundell, T. Lancaster, I. Franke, W. Hayes, F.L. Pratt, L. Bohatý, and P. Becker, Phys. Rev. B **81**, 214403 (2010).

[17] G. Nénert, M. Isobe, C. Ritter, O. Isnard, A.N. Vasiliev, and Y. Ueda, Phys. Rev. B **79**, 064416 (2009).

[18] G. Nénert, M. Isobe, I. Kim, C. Ritter, C.V. Colin, A.N. Vasiliev, K.H. Kim, and Y. Ueda, Phys. Rev. B **82**, 024429 (2010).

[19] O. Janson, G. Nénert, M. Isobe, Y. Skourski, Y. Ueda, H. Rosner, and A.A. Tsirlin, Phys. Rev. B **90**, 214424 (2014).

[20] L. Ding, C. V. Colin, C. Darie, J. Robert, F. Gay, and P. Bordet, Phys. Rev. B **93**, 064423 (2016).

[21] S. V. Streltsov and D.I. Khomskii, Phys. Rev. B **77**, 064405 (2008).

[22] L. Ding, C. V. Colin, C. Darie, and P. Bordet, J. Mater. Chem. C **4**, 4236 (2016).

[23] J. Rodriguez-carvajal, Phys. B **192**, 55 (1993).

[24] P.J. Becker and P. Coppens, Acta Crystallogr. Sect. A **30**, 129 (1974).

[25] H.T. Stokes, B.J. Campbell, and D.M. Hatch, *ISOTROPY Software Suite, Iso.Byu.Edu* (n.d.).

[26] J.M. Perez-Mato, S.V. V Gallego, E.S.S. Tasci, L. Elcoro, G. De Flor, M.I.I. Aroyo, G. de la Flor, and M.I.I. Aroyo, Annu. Rev. Mater. Res. **45**, 217 (2015).

[27] W.H. Baur, Acta Crystallogr. Sect. B Struct. Crystallogr. Cryst. Chem. **30**, 1195 (1974).

[28] J.M. Perez-Mato, J.L. Ribeiro, V. Petricek, and M.I. Aroyo, J. Phys. Condens. Matter **24**, 163201 (2012).

[29] In the spin context Isotropy means "the same in all directions" in "spin-space". That is, an isotropic exchange interaction is one where the x-components of the two spins interact the same as the y-components, and both interact the same as the z-components. For two spin, the isotropic interaction is the Heisenberg interaction ($\vec{S}.\vec{S}$).

[30] T.A. Kaplan and N. Menyuk, Philos. Mag. **87**, 3711 (2007).

[31] G. Nénert, C. Ritter, M. Isobe, O. Isnard, a. Vasiliev, and Y. Ueda, Phys. Rev. B **80**, 024402 (2009).

[32] G. Nénert, I. Kim, M. Isobe, C. Ritter, A.N. Vasiliev, K.H. Kim, and Y. Ueda, Phys. Rev. B **81**, 184408 (2010).

[33] G.J. Redhammer, G. Roth, W. Treutmann, W. Paulus, G. André, C. Pietzonka, and G. Amthauer, J. Solid State Chem. **181**, 3163 (2008).

[34] S. Petit, Collect. SFN **12**, 105 (2011).

[35] F. V. Temnikov, E. V. Komleva, Z. V. Pchelkina, and S. V. Streltsov, JETP Lett. https://doi.org/10.1134/S0021364019210033 (2019).

[36] E. Ressouche, M. Loire, V. Simonet, R. Ballou, A. Stunault, and A. Wildes, Phys. Rev. B - Condens. Matter Mater. Phys. **82**, 100408(R) (2010).

[37] M. Mostovoy, Phys. Rev. Lett. **96**, 067601 (2006).

[38] N. Terada, D.D. Khalyavin, P. Manuel, Y. Tsujimoto, K. Knight, P.G. Radaelli, H.S. Suzuki, and H. Kitazawa, Phys. Rev. Lett. **109**, 097203 (2012).

[39] J.S. White, C. Niedermayer, G. Gasparovic, C. Broholm, J.M.S. Park, A.Y. Shapiro, L.A. Demianets, and M. Kenzelmann, Phys. Rev. B - Condens. Matter Mater. Phys. **88**, 060409(R) (2013).